\documentclass[prl,aps,twocolumn]{revtex4}
\usepackage{graphicx}
\usepackage{latexsym}
\begin{document}
\title{Cosmological  consequences of generalised RS II braneworlds}
\author{Supratik Pal \footnote{Electronic address: {\em supratik\_v@isical.ac.in}}}
${}^{}$
\affiliation{Physics and Applied Mathematics Unit, 
Indian Statistical Institute, 
203 B.T.Road, Kolkata 700 108, India}
\newcommand{\be}{\begin{equation}}
\newcommand{\ee}{\end{equation}}
\newcommand{\bea}{\begin{eqnarray}}
\newcommand{\eea}{\end{eqnarray}}
\newcommand{\bml}{\begin{subequations}}
\newcommand{\eml}{\end{subequations}}
\newcommand{\bfig}{\begin{figure}}
\newcommand{\efig}{\end{figure}}

\vspace{.5in}

\begin{abstract}

We discuss certain features of cosmology in a generalised RS II braneworld scenario. In this scenario, the bulk is given by a Schwarzschild-anti de Sitter or a  Vaidya-anti de Sitter black hole
in which an FRW brane is consistently embedded,  
resulting in  modifications of the 4-dimensional Friedmann equations. 
We analyse how the scenario can be visualised and
discuss the significance of each term in these modified equations both for early time and for
late time cosmology. 
We further analyse the perturbation equations, based on Newtonian as well as relativistic
perturbations and show that the scenario
has the potentiality to explain structure formation by the ``Weyl fluid''
arising from embedding geometry. 
The results thus obtained are confronted with observations as well.

\end{abstract}


\maketitle

\section {Introduction}

The Randall-Sundrum braneworld model \cite{rs} provides us with a unique mechanism of
solving the hierarchy problem and obtaining an effective 
4-dimensional Newton's law of  gravity valid at large as well as small length scales,
by the method of embedding our 4D universe in a higher dimensional anti-de Sitter (AdS$_5$) space.
 Of late, a more general, and
geometrically rich, picture of obtaining an effective 4D theory of gravity, 
by projecting the bulk field equations onto the brane,
has been formulated by Shiromizu et al \cite{eee}.  
In this scenario, the bulk spacetime is not necessarily AdS$_5$, rather
a generalised version of it. 
The so-called \textit{effective Einstein equation on the brane},  thus obtained,
 not only gives important insight to the bulk-brane
interplay but also raises the possibility of explaining the gravitational phenomena of the 4D universe
from a  broader perspective. Since the formalism was brought forth, it has been applied
to explain several gravitational aspects,
leading to interesting results which justify further investigation on
the prospects of braneworlds.
A thorough discussion of these issues is available in the review of S. Kar \cite{skrev} in this volume. 

The cosmological aspects of this theory were developed at a rather later stage, due to
the complicated equations arising from the extra terms involved in the theory,
which makes the theory difficult to probe at a first go.
The basic aim of this article is to provide a brief review of what has been
done in the cosmological context of braneworlds so far and show that, in spite of
great complications involved, the cosmological consequences of this scenario
results in interesting physics which is worth investigating further.


\section{FRW brane, Sch-AdS bulk}

In the braneworld scenario,
the bulk may be either empty with only a bulk cosmological constant,
or it may consist of non-standard model fields minimally or non-minimally
coupled to gravity or to brane matter.
For empty bulk, the bulk metric in which an 
FRW brane can be consistently embedded, is given by a 5-dimensional Schwarzschild-Anti de Sitter (Sch-AdS$_{5}$)
black hole  \cite{maartrev,scads5} 
\be
d S_5^2 = - f(r) dt^2 + \frac{dr^2}{f(r)} + r^2 d \Sigma_3^2 
\ee
where $\Sigma_3$ is the 3-space and the function $f(r)$ is given by
\begin{equation}
f(r) =  k  - \frac{\Lambda_5}{6} r^2 - \frac{m}{r^2}
\end{equation}
where $k (= 0, \pm 1)$ is the spatial curvature. Here $m$ is a constant which is
identical to the 5D analogue of the Schwarzschild mass of the bulk black hole. 

An important point to note here is that
the bulk metric presented here is asymptotically anti-de Sitter.
 When the mass of the 5D black hole vanishes, this metric can be recast in the 
warped RS II form by coordinate transformation. Therefore, the scenario is a generalisation
of the RS II braneworlds, which reflects cosmology on the brane.  In this sense,
this setup may be called {\em generalised RS II braneworlds}.

The Sch-AdS$_5$ black hole provides a novel way of visualising cosmological phenomena on the 4D universe.
In this scenario, the brane is moving in the bulk, with its radial trajectory given by
$r(\tau)$, where $\tau$ is  the proper time on the brane.  
With $u^{\mu} = (\dot t, ~\dot r)$ the  velocity vector on the brane, 
the normalisation condition is given by
\be
g_{\mu \nu} u^{\mu} u^{\nu} = -f \dot t^2 + \frac{\dot r^2}{f} = -1
\ee
and the unit normal vector $n^{\mu}$ , satisfying the orthonormality conditions,  is given by
\be
n^{\mu} = (\dot r, ~-\frac{\sqrt{f + \dot r^2}}{f})
\ee
With this, we can get the induced metric on the brane 
\be
ds^2 = - d\tau^2 + r^2(\tau) d\Omega_3^2
\ee
which is, clearly, an FRW metric on the brane, with the scale factor $a(\tau)$  being identified with the
radial trajectory $r(\tau)$.
So, the interpretation here is as follows:
What we call an expanding universe is visualised by a brane-based
observer only. For a bulk-based observer located somewhere outside the bulk black hole horizon,
this expansion is exactly identical to the movement of the brane along the radial trajectory
of the black hole. 


Given this situation, the important question is: How do the Friedmann equations on the
brane look like?
The review of S. Kar in this volume \cite{skrev} discusses in details that  the 4D Einstein
equation on the brane is a generalisation of the standard Einstein equation. The so-called
{\em Effective Einstein Equation} is given by
\cite{eee}
\begin{equation}
G_{\mu \nu} = -\Lambda g_{\mu \nu} + \kappa_4^2 T_{\mu \nu} +
\kappa_5^4 {\cal S}_{\mu \nu} -{\cal E}_{\mu \nu}
\label{eee}
\end{equation}
where the 4D cosmological constant and coupling constant are related
to their 5D counterparts by 
\be
\Lambda= \frac{\kappa_5^2}{2} \left(\Lambda_5 + 
\frac{\kappa_5^4 \lambda_b^2}{6} \right) 
\hspace{.5 cm}, \hspace{.5 cm} 
\kappa_4^2 = \frac{\kappa_5^2 \lambda_b}{6}
\ee 
${\cal S}_{\mu \nu}$ is the quadratic contribution from brane energy-momentum tensor
\be
{\cal S}_{\mu \nu} = \frac{1}{12} T T_{\mu \nu} - \frac{1}{4} T_{\mu
  \alpha} T^\alpha_\nu + \frac{1}{8} T_{\alpha \beta} T^{\alpha \beta}
g_{\mu \nu} -\frac{1}{24} T^2 g_{\mu \nu}
\ee 
and ${\cal E}_{\mu \nu}$ is the bulk Weyl tensor, projected
onto the brane, which is given by 
\be
{\cal E}_{\mu \nu} = C_{ABCD} n^C n^D g_\mu^A g_\nu^B
\label{weyl}
\ee
Thus, along with the usual contributions from the cosmological constant and matter on the 4D universe,
the EEE contains two additional terms : 
\begin{itemize}
\item ${\cal S}_{\mu\nu}$ :  the local correction 
from brane matter

\item ${\cal E}_{\mu\nu}$ : the nonlocal correction from bulk geometry.
\end{itemize}

 For a perfect fluid on the brane, using the symmetry properties,
these braneworld corrections can  be added up to the usual brane
energy-momentum tensor. The effective 
energy density, pressure, momentum density and 
anisotropic stress thus obtained are, respectively, given by \cite{maartrev}:
\bea
\rho^{\text{eff}} &=& \rho +\frac{\rho^2}{2\lambda_b} + \rho^* 
\label{cos1} \\ 
p^{\text{eff}} &=& p  + \frac{\rho}{2\lambda_b} (\rho +2p)+\frac{\rho^*}{3} 
\label{cos2} \\
q_\mu ^{\text{eff}} &=& q^*_\mu \\
\pi_{\mu\nu}^{\text{eff}} &=& \pi^*_{\mu\nu}
\eea
Further, 
the brane matter conservation equation 
\be
\dot\rho + 3 \frac{\dot a}{a} (\rho + p) = 0
\ee
leads to  the conservation equation for the Weyl term (via 4D Bianchi identity on the brane
$\nabla^\mu G_{\mu\nu}=0$)
\be
\dot\rho^* + 4 \frac{\dot a}{a} \rho^* = 0
\label{cos3}
\ee
Hence the Weyl density $\rho^*$ behaves as
\be
\rho^* = \frac{C}{a^4}
\label{cos4}
\ee
where $C$ is a constant which is basically the rescaled bulk black hole mass.

Further, for a bulk compatible to FRW geometry on the brane,
$q^*_\mu = 0 = \pi^*_{\mu\nu}$. With the use of the effective Einstein equation (\ref{eee})
and the effective quantities obtained in equations (\ref{cos1}) and (\ref{cos2})
 for a perfect fluid on the brane,
the Friedmann equation and the covariant Raychaudhuri equation turn out to be
\bea
H^2 &=& \frac{\kappa_4^2}{3} \rho^{\text{eff}} + \frac{\Lambda}{3} -
\frac{k}{a^2}  \label{cos6} \\
\dot H &=&-\frac{\kappa_4^2}{2} (\rho^{\text{eff}} + p^{\text{eff}}) +
\frac{k}{a^2}
\label{ecos7} 
\eea
Obviously, these equations are quite different from the standard 4D equations
due to the presence of the quadratic terms and the Weyl terms.

Let us now explore the roles of these additional terms on 4D cosmology.
The quadratic term ${\cal S}_{\mu \nu}$ plays a significant 
role in the early universe ($\rho^2
\gg \lambda_b$).  
For example,  this   leads to a faster
  Hubble expansion at high energies and a more strongly damped
  evolution of the inflaton field \cite{chaoinfl}. Thus the brane universe inflates at a
  much faster rate than what is expected from standard cosmology.
Further, with this term, the modified
  equations make it possible to explain the inflationary scenario
 which is driven not by a 
  inflaton field on the brane ({\em i.e.}, not by any 4D field as
  required in standard cosmology) but by a dilaton
  field in the bulk \cite{infl}. 
However, the quadratic term  is negligibly small at late times as
$\rho^2 \ll \lambda_b > (100 GeV)^4$.
Hence, its role is relevant so far as only the early universe is concerned.

On the other hand, the role of ${\cal E}_{\mu \nu}$ is to supply an additional
perfect fluidlike effect to the actual on-brane perfect fluid
(since, from the previous arguments, $q^*_\mu = 0 = \pi^*_{\mu\nu}$).
 The so-called {\em Weyl fluid} arises as the tidal  effect of the
5D black hole in the bulk and its density contribution $\rho^*$ is
related to the mass of the bulk black hole.
Hence,  in order to have a realistic contribution
from the Weyl term on the brane, we need the black hole mass to be
positive, so that $\rho^* > 0$. 
Since for a vacuum bulk, Eq (\ref{cos4}) reveals that the Weyl fluid is strictly radiation-like,
it does not play any significant role in late
time cosmologies. It can, at best, slightly modify the standard
perturbative analysis. Its role has been extensively studied  for
metric-based perturbations \cite{pertmet}, density perturbations on
large scales \cite{pertden}, curvature perturbations  \cite{pertcur} and the
Sachs-Wolfe effect \cite{brsachs}, vector perturbations
\cite{pertvec}, tensor perturbations \cite{pertten} and CMB
anisotropies \cite{pertcmb}. In all the cases, the effect has been
found to be slightly enhanced from the standard analysis. 
However, we shall show in the subsequent discussions that the Weyl
fluid can have a crucial role in late time cosmologies as well when the bulk is
not necessarily empty.


\section{FRW brane, Vaidya-AdS bulk}

Till now we have discussed brane cosmology with  empty bulk.
However, as already mentioned, the bulk may, in principle contain non-standard model fields,
for which the above situation is further generalised.
In this case, the bulk metric,
for which the FRW geometry on the brane is recovered, is given by a radiative
5D Vaidya anti-de-Sitter 
black hole (VAdS$_5$).
In terms of transformed (null)
coordinate $v = t + \int d r/f$, the bulk metric can be written as
\begin{equation}
d S_5^2 = - f(r, ~v) ~dv^2 + 2 dr ~dv + r^2 d \Sigma_3^2 
\label{eqd1} 
\end{equation}
where $\Sigma_3$ is the 3-space. For a spatially flat brane, the function $f(r, ~v)$ is given by
\begin{equation}
f(r, ~v) =  \frac{r^2}{l^2} - \frac{m(v)}{r^2}
\label{eqd2} 
\end{equation}
with the length scale $l$ related to the bulk (negative) cosmological constant
by $\Lambda_5 = -6 / l^2$ and
$m(v)$ is the variable mass of the Vaidya black hole.

In the case of such a radiative bulk,
 the effective  Einstein equation on the brane is further modified to \cite{maartbulk, lang2}
\begin{equation}
G_{\mu \nu} = -\Lambda g_{\mu \nu} + \kappa_4^2 T_{\mu \nu} +
\kappa_5^4 {\cal S}_{\mu \nu} -{\cal E}_{\mu \nu} + {\cal F}_{\mu \nu}
\end{equation}

Clearly, in this scenario, the extra terms come from three sectors : 
\begin{itemize}
 \item ${\cal S}_{\mu \nu}$ : a quadratic term
from brane energy-momentum tensor
\item ${\cal E}_{\mu \nu}$ : a geometric term from the bulk Weyl tensor 
projected onto the brane
\item ${\cal F}_{\mu \nu}$ : a term involving the brane-projection of the 
bulk energy-density. 
\end{itemize}
The combined effect
of the last two terms is related to the sumtotal of the  mass of bulk black hole and the radiation field,
and this is now  the {\em Weyl fluid} that supplies an additional perfect fluid-like
effect to the usual brane perfect fluid.

Here the exclusive contribution from the bulk matter is given by the energy-momentum tensor of a null dust
\be
T_{AB}^{\text{bulk}} = \psi ~q_A q_B
\label{vaidmat} 
\ee
where $q_A$ are now the ingoing null vectors and $\psi \propto dm/dv$ is the
rate of ingoing radial flow to the bulk black hole, $m(v)$ being the
resultant of the masses of VAdS$_5$ black hole and the radiation field.
This type of bulk  can exchange energy with the brane along the radial direction.
There are extensive study in the literature on the energy-exchange between bulk and brane.
(see, for example,  
\cite{lang2, maartbulk}). 
Consequently, the brane matter conservation equation is modified to 
\be
\dot\rho + 3 \frac{\dot a}{a} (\rho + p) = -2 \psi
\label{vaidden} 
\ee
which now contains an effect of bulk matter onto the brane and guarantees the
null flow from the radiative bulk black hole. 
Also, the Bianchi identity on the brane $\nabla^\mu G_{\mu\nu}=0$ leads to
the equation governing the evolution of the Weyl fluid 
\be
\dot\rho^* + 4 \frac{\dot a}{a} \rho^* = 2 \psi - \frac{2 \kappa_5^2}
{3 \kappa_4^2} \left[\dot \psi + 3 \frac{\dot a}{a} \psi \right] 
\label{vaidweyl1} 
\ee
The difference of the above  equation
with Eq (\ref{cos3}) for empty bulk is remarkable. 
For a bulk with matter, the coupling term involving $\psi$ 
determines the nature of the Weyl fluid.
A general expression for  this is now given by \cite{lang2}
\be
\rho^* = \frac{C(\tau)}{a^4}
\label{vaidweyl2} 
\ee
$\tau$ being the proper time on the brane.
Clearly, when the bulk is not strictly matter-free, then the Weyl parameter $C(\tau)$ is
no longer a constant, and consequently, $\rho^*$ no longer 
behaves like radiation. It is only when the bulk is empty, we get back
the radiation-like behaviour of the Weyl fluid from Eq (\ref{vaidweyl1}).

It is obvious from the preceding discussion  that the Vaidya-AdS bulk
scenario is so far the most generalised description of the braneworlds 
for which cosmology of the 4D world is relevant. Thus, this is the most general
{\em generalised RS II braneworld} scenario.

The components of the effective energy-momentum tensor $T^{\text{eff}}_{\mu \nu}$ are now
given by \cite{sustr}
\bea
\rho^{\text{eff}} &=& \rho +\frac{\rho^2}{2\lambda_b} + \frac{C(\tau)}{a^4} 
\label{eqb15} \\ 
p^{\text{eff}} &=& p  + \frac{\rho}{2\lambda_b} (\rho +2p)+\frac{C(\tau)}{3 a^4}
\label{eqb16}  
\eea
As before, for a Vaidya-Ad$S_5$ bulk compatible to FRW geometry on the brane,
both $q^{\text{eff }}_\mu$ and $\pi^{\text{eff }}_{\mu\nu}$  vanish. We are thus
left with a perfect fluid like effect on the brane, constituted of brane and 
bulk matter-energy and bulk geometry, with an evolving Weyl fluid.

The final modifications to 
the Friedmann equation and the covariant Raychaudhuri equation on the brane are
due to the collective effect of the embedding geometry, brane matter and  bulk matter.
In terms of the brane and bulk quantities, 
these generalised equations read \cite{sucol}
\bea
H^2 &=& \frac{\kappa_4^2}{3} \rho^{\rm eff}
 + \frac{\Lambda}{3} - \frac{k}{a^2} \\
\dot H &=& -\frac{\kappa_4^2}{2} \left(\rho^{\rm eff} + p^{\rm eff} \right) + \frac{k}{a^2} - \frac{\kappa_5^2}{3} \psi
\eea

It is worth mentioning that these  equations are 
radically different from those for empty bulk due to the presence of
an evolving Weyl parameter $C(\tau)$. So, they should, in principle,
give rise to new physics on the brane.


\section{``Newtonian'' perturbations}

Given the modified Friedmann equations, we will now engage ourselves in studying cosmological
perturbations on the brane. 
Following  Newtonian analysis of perturbations from
gravitational instability,  we will demonstrate that 
the  growth of perturbations of the Weyl fluid can
take care of the fluctuations required by the inhomogeneous
 local  universe, and thus, have the potentiality to explain structure formation 
without the need for dark matter \cite{sustr}.


The equations
of hydrodynamics now  involve the quadratic brane correction and the Weyl fluid
correction to the brane perfect fluid. In terms of the effective quantities, these equations are given by
\bea
\frac{\partial \rho^{\text{eff}}}{\partial t} + \overrightarrow\nabla. 
(\rho^{\text{eff}} ~\overrightarrow v^{\text{eff}})&=&0 
\label{eqc1}  \\
\frac{\partial \overrightarrow v^{\text{eff}}}{\partial t} + 
(\overrightarrow v^{\text{eff}}.\overrightarrow\nabla)
\overrightarrow v^{\text{eff}}&=&- \frac{\overrightarrow\nabla 
p^{\text{eff}}}{\rho^{\text{eff}}} -\overrightarrow\nabla \Phi^{\text{eff}} 
\label{eqc2}   \\
\nabla^2 \Phi^{\text{eff}}&=&4 \pi G \rho^{\text{eff}} 
\label{eqc3}  
\eea
where $\overrightarrow v^{\text{eff}}$ is the velocity
field in the {\em effective} perfect fluid.
It should be noted that the term $\Phi^{\text{eff}}$ is not the usual
Newtonian potential but the effective gravitational potential
which is the resultant effect of the brane and bulk parameters in the form
of effective quantities.

Let us now consider small perturbation about the initial unperturbed effective quantities.
Perturbation on the effective density (the so-called
 {\em effective density contrast}) is given by
\be
\rho^{\text{eff}}(\overrightarrow x, ~\tau) = \bar\rho^{\text{eff}}(\tau) 
(1 + \delta^{\text{eff}}(\overrightarrow x, ~\tau)) 
\label{eqc4} 
\ee
whereas the perturbation in the effective gravitational potential is 
\be
 \Phi^{\text{eff}}(\overrightarrow x, ~\tau) = \Phi_0^{\text{eff}} + \phi^{\text{eff}}
 \label{eqc5} 
\ee
where $\bar\rho^{\text{eff}}(\tau)$ and $\Phi_0^{\text{eff}}$ are respectively the
unperturbed effective density and effective  
potential and $\delta^{\text{eff}}$
and $\phi^{\text{eff}}$ are their corresponding fluctuations. 
For completion, we mention here that a perturbation in the Weyl fluid would mean
a perturbation on the bulk geometry.
There is a possibility that this may destabilise the brane itself. In this context, however,
we assume that the brane remains stable even after such perturbations, which can only be
guaranteed if one analyses the effects of perturbations on the full 5D bulk metric.

Expressing in terms of comoving coordinates
$v^{\text{eff}} = \dot a ~r+  u^{\text{eff}}$
and neglecting terms of second or higher order in the equations we arrive at the
simplified perturbation equations 
\bea
\frac{\partial \delta^{\text{eff}}}{\partial \tau} + \frac{1}{a}\overrightarrow\nabla_r.
\overrightarrow u^{\text{eff}}&=&0 
\label{eqc6} \\
\frac{\partial \overrightarrow u^{\text{eff}}}{\partial \tau} + \frac{\dot a}{a} 
\overrightarrow u^{\text{eff}}
&=&  - \frac{1}{a} \frac{\overrightarrow\nabla_r p^{\text{eff}}}{\bar \rho^{\text{eff}}}
- \frac{1}{a} \overrightarrow\nabla_r \phi^{\text{eff}} 
\label{eqc7} \\
\nabla_r^2 \phi^{\text{eff}}&=&4 \pi G a^2 \bar\rho^{\text{eff}} \delta^{\text{eff}}
\label{eqc8} 
\eea
The above set of equations have a unique solution 
given by 
\be
\delta^{\text{eff}}(\overrightarrow x, ~\tau) = \sum \delta_k^{\text{eff}}(\tau) 
~e^{i \overrightarrow k. \overrightarrow x} 
\label{eqc9} 
\ee
From now on, we shall express the perturbations in terms of 
 the inverse Fourier transform of the above: 
\be
\delta_k^{\text{eff}}(\tau) = \frac{1}{V} \int \delta^{\text{eff}}
(\overrightarrow x, ~\tau) ~e^{- i \overrightarrow k. \overrightarrow x}
~d^3 \overrightarrow x
\label{eqc10} 
\ee
For a barotropic fluid, the effective pressure is a function of the
effective density only. Hence,  equations (\ref{eqc6}) - (\ref{eqc8}),
with the Fourier mode solution, 
transform into a linear second order differential equation for the perturbation 
$\delta_k^{\text{eff}}$ 
\be
\frac{d^2 \delta_k^{\text{eff}}}{d \tau^2} + 2\frac{\dot a}{a}
\frac{d \delta_k^{\text{eff}}}{d \tau} - \left[4 \pi G \bar\rho^{\text{eff}}
- \left(\frac{c^2_{\text{eff}} k}{a} \right)^2\right]\delta_k^{\text{eff}} = 0
\label{eqc11} 
\ee
where $c^2_{\text{eff}}$ is  the square of the effective sound speed which is given by
\bea
c^2_{\text{eff}} &=& \frac{\dot p^{\text{eff}}}{\dot\rho^{\text{eff}}}
= \left[c_s^2 + \frac{\rho + p}{\rho + \lambda_b} 
+ \frac{4 \rho^*}{9 (\rho + p)(1 + \rho/\lambda_b)}\right] \nonumber \\
{} &\times& \left[1 + \frac{4 \rho^*}{3 (\rho + p)(1 + \rho/\lambda_b)} \right]^{-1} 
\eea


The main job now is to analyse the second order differential equation (\ref{eqc11})
for the effective perturbation.
The perturbations of the effective fluid will grow at late times and 
will account for the required amount of gravitational instability only if the Weyl density 
 redshifts more slowly than baryonic matter density, so that  it can 
dominate over  matter at late times. 
Recall that for a Sch-AdS bulk (empty bulk), the Weyl fluid is strictly radiation-like, hence
it redshifts at a faster rate than baryonic matter, eventually being negligible at late times. 
But for the VAdS bulk (radiative bulk), 
 the nature and evolution of the Weyl fluid is governed
by Eq (\ref{vaidweyl1}). 
This equation can be written conveniently as
\begin{equation}
\dot\rho^* + 4 \frac{\dot a}{a} \rho^* = {\cal{Q}}
\label{eqc14} 
\end{equation}
where ${\cal{Q}}$ is a coupling term which is determined 
by the projected bulk energy density $\psi$. 
Since the bulk informations are not known {\em ab initio}, this coupling term is given by
a physically reasonable
and consistent ansatz. Certain earlier attempt of choosing an ansatz are listed in
\cite{sucol, ansatz, excos4}. We note here that,
in order to study the evolution of the Weyl fluid as function of the scale factor,
the general ansatz for the coupling term should be of the form \cite{sustr}
\be
{\cal{Q}} = \alpha H \rho^*
\label{ansatz}
\ee
with $\alpha > 0$.  For this type of ansatz, the Weyl fluid behaves as
\be
\rho^* \propto \frac{1}{a^{(4 - \alpha)}}
\label{eqc16} 
\ee
Consequently, Eq (\ref{vaidweyl2}) reveals that 
the Weyl parameter is given by  
\be
C(\tau) = C_0 ~a^{\alpha}(\tau)
\ee
 where $C_0$ is its initial value at the matter-dominated epoch.
Note that the Weyl fluid is strictly radiation-like only if
$\alpha = 0$, for which we recover the Sch-AdS bulk scenario, which is a special case 
of this generalised description.
So, in general, the nature of the Weyl fluid depends on
the coupling strength $\alpha$. In order that the Weyl fluid dominates over matter 
the coupling strength $\alpha$ should have a value within the range $1 < \alpha < 4$,
as obtained from this ``Newtonian'' analysis. Later, we shall put  more stringent bounds
on the value of this parameter from relativistic analysis as well as from observational
ground.


Let us now get  back to the the perturbation
equation (\ref{eqc11}) of the effective density which is a sum-total
of three quantities. As discussed before, the quadratic term being negligible at late times,
 the effective density 
at late times is practically given by
\begin{equation}
\rho^{\text{eff}} \approx \rho^{(b)} + \rho^* 
\label{eqc12} 
\end{equation}
where $\rho^{(b)}$ is the baryonic density.
Thus, along with the usual matter
density, here we have an additional (Weyl) density contributing to the
total density that governs the perturbation equation.
This Weyl density, being geometric, is essentially non-baryonic. Consequently,
we decompose Eq (\ref{eqc11}) by separating the baryonic (matter) part from the non-baryonic (Weyl) part, which results in two separate equations, one each for each of the fluids 
\begin{eqnarray}
\frac{d^2 \delta^{(b)}}{d \tau^2} + 2\frac{\dot a}{a} \frac{d \delta^{(b)}}{d \tau} 
= 4 \pi G \bar\rho^{(b)} \delta^{(b)} + 4 \pi G \bar\rho^* \delta^*
\label{eqc18} \\
\frac{d^2 \delta^*}{d \tau^2} + 2\frac{\dot a}{a} \frac{d \delta^*}{d \tau} 
= 4 \pi G \bar\rho^* \delta^* + 4 \pi G \bar\rho^{(b)}  \delta^{(b)} 
\label{eqc19} 
\end{eqnarray}
where $\delta^{(b)}$ and $\delta^*$ are the fluctuations of baryonic
matter and Weyl fluid respectively. 
In the above equations, the term involving sound speed has been neglected as we are interested only in the growing mode fluctuations.  
With $\Omega^{(b)}  \ll \Omega^*$, the relevant growing mode solution for Eq (\ref{eqc19})
can be expressed  
as a function of the redshift as
\begin{equation}
\delta^*(z) = \delta^*(0) (1 + z)^{-1}
\label{eqc20} 
\end{equation}
which, when put back into the fluctuation equation (\ref{eqc18})
of baryonic density,  gives 
\begin{equation}
\frac{d^2 \delta^{(b)}}{d \tau^2} + 2\frac{\dot a}{a}
\frac{d \delta^{(b)}}{d \tau} = 4 \pi G \bar\rho^* \delta^*(0) (1 + z)^{-1}
\label{eqc21} 
\end{equation}
Since the late time behaviour of the expansion of the universe
in RS II is the same as the standard cosmological solution for the 
scale factor \cite{maartbulk, maart6},
for a spatially flat ($k = 0$) brane, we have the scale factor at late time
\begin{equation}
a(\tau) = \left(\frac{3}{2} H_0 \tau\right)^{2/3(w + 1)} 
\label{eqc13} 
\end{equation}
Considering $\Omega^{\rm tot} \approx \Omega^* \approx 1$ at present time, 
Eq (\ref{eqc21}) simplifies to
\begin{equation}
a^{3/2} \frac{d}{d a} \left( a^{-1/2} \frac{d \delta^{(b)}}{d a}\right) 
+ 2 \frac{d \delta^{(b)}}{d a} = \frac{3}{2} \delta^*(0)
\label{eqc22} 
\end{equation}
which readily gives a solution of the form  \cite{sustr}
\begin{equation}
\delta^{(b)}(z) = \delta^*(z) \left(1 - \frac{1+z}{1+z_N}\right) 
\label{eqc23} 
\end{equation}
where  the scale factor is related to the redshift function by $a \propto (1+z)^{-1}$.

 Eq (\ref{eqc23}) reveals that at a redshift close to $ z_N$, 
the baryonic fluctuation $\delta^{(b)}$ almost vanishes but the Weyl fluctuation 
$\delta^*$ still remains finite. This implies that even if the baryonic fluctuation 
is very small at a redshift of $z_N \approx 1000$,
as confirmed by CMB data \cite{cmb}, the fluctuations of the Weyl fluid 
had a finite amplitude during that time.  
On the other hand, at a redshift much less than  $z_N$
the baryonic matter fluctuations are of equal amplitude as
the Weyl fluid fluctuations. This is precisely what is required to
explain the formation of structures we see today.

The bottomline here is that in this analysis, we get a result which
resembles  standard
cosmological relation, but has an essential distinction that  nowhere we have to introduce
dark matter by hand. Rather, 
here the fluctuations are governed by the Weyl density  fluctuation $\delta^*$ which
is a product of the embedding geometry  via a modified Einstein equation
 in the  braneworld scenario. 

Another  point is worth mentioning here.  
The \textit{effective} equation of state parameter 
\begin{equation}
w^{\text{eff}} = \frac{p^{\text{eff}}}{ \rho^{\text{eff}}} 
= \frac{p^{(b)}  + \rho^{(b)}  (\rho^{(b)} +2p^{(b)})/ 2\lambda_b + C(\tau)/ 3 a^4}
{\rho^{(b)} +\rho^{(b)2}/ 2\lambda_b + C(\tau)/ a^4}
\end{equation}
 reduces to (in this context of matter-dominated era)
\begin{equation}
w^{\text{eff}} \approx \frac{1}{3 (1+ C a^{1 - \alpha})}
\label{eos}
\end{equation}
which bears significant difference from the equation of state of cold dark 
matter ($w = 0$). Thus, the theory gives rise to a geometric fluid which is very
different from dark matter in origin and nature but has the potentiality
to play the role of dark matter in cosmological context.


\section{Relativistic perturbations}

Let us now proceed further to study relativistic perturbations on the brane
based on the above analysis. The scenario, in brief, is as follows:
 Here the cosmological dynamics is governed by a  two-fluid system,
out of which one is the material fluid  $\rho^{(b)}$, which is the ordinary (baryonic) matter, 
and another a geometric fluid $\rho^{*}$, termed as the Weyl fluid.
These two fluids interact and exchange energy between them  in such a way that the
total (effective) density on the brane is given by $ \rho = \rho^{(b)} + \rho^{*}$.
\cite{suden}
  
Given this scenario, the 
comoving fractional gradients of density and expansion are expressed, following
usual general relativity, as  
\begin{eqnarray}
\Delta^{(i)}_{\mu} &=& \frac{a}{\rho^{(i)}} D_{\mu} \rho^{(i)}\\
Z_{\mu} &=& a D_{\mu} \Theta \\
\Delta_{\mu} &=& \frac{a}{\rho} D_{\mu} \rho
\end{eqnarray}

Further, the 
conservation equations (\ref{vaidden}) and (\ref{vaidweyl1}) can be  written composedly as
\begin{equation}
\dot\rho^{(i)} + \Theta (\rho^{(i)} + p^{(i)}) = I^{(i)}
\end{equation}
where a superscript $(i)$ denotes the quantities for the $i$-th fluid and $I^{(i)}$ is the corresponding interaction term. Written explicitly in the bulk-brane scenario discussed here,
the interaction terms are:
\begin{eqnarray}
I^{(b)} &=& - 2 \psi \\
I^{*} &=& 2 \psi - \frac{2}{3} \left ( \frac{\kappa_{5}}{\kappa} \right )^{2} 
\left( \dot \psi + 3 \frac{\dot a}{a} \psi \right) 
\end{eqnarray}

In what follows we shall restrict ourselves to the discussion of 
the  Einstein-de Sitter brane universe for which $\Omega^* = 1, \Omega_\Lambda = 0$.
Here, with the above notations, the linearised evolution equations turn out to be

\begin{eqnarray}
\dot\Delta^{(i)}_{\mu}  & = & \left ( 3 H w^{(i)} -  \frac{I^{(i)}}{\rho^{(i)}}     \right ) \Delta_{\mu}^{(i)} - (1 + w^{(i)}) Z_{\mu}  \nonumber\\
& - & \frac{c^{2}_{s} ~ I^{(i)} }{\rho^{(i)} (1 + w)}  \Delta_{\mu} - \frac{3 a H I^{(i)}_{\mu}}{\rho^{(i)}}  + \frac{a}{\rho^{(i)}} D_{\mu} I^{(i)}  \\
\dot Z_{\mu} & + & 2 H Z_{\mu}  =  -  \frac{\kappa^{2}}{2} \rho \Delta - \frac{c^{2}_{s}}{1 + w} D_{\mu} D^{\nu} \Delta_{\nu}  \nonumber\\
& + &
\frac{\kappa^{2}_{_{5}} \psi}{1 + w}  c^{2}_{s} \Delta_{\mu} - a \kappa^{2}_{_{5}} D_{\mu} \psi
\end{eqnarray}

We shall now try to express the above equations in terms of covariant quantities. 
These  density perturbations are governed by the 
fluctuation of the following covariant projections 
\begin{equation}
\Delta^{(i)} = a~ D^{\mu} \Delta^{(i)}_{\mu} , \Delta = a~ D^{\mu} \Delta_{\mu} ,  Z = a~ D^{\mu}Z_{\mu}
\end{equation}

The covariant density perturbation equations on the brane, when expressed in terms of
the above covariant quantities, turn out to be 
\begin{eqnarray}
\dot\Delta^{(i)} & = & \left ( 3 H w^{(i)} - \frac{I^{(i)}}{\rho_{(i)}} \right ) \Delta^{(i)} - (1 + w^{(i)}) Z  \nonumber\\
& - &
\frac{c^{2}_{s} ~ I^{(i)} }{\rho^{(i)} (1 + w)}  \Delta - \frac{3 a^{2} H D^{\mu} I^{(i)}_{\mu}}{\rho^{(i)}}  + \frac{a^{2}}{\rho^{(i)}} D^{2} I^{(i)} \\
 \dot Z & + & 2HZ = - \frac{\kappa^{2}}{2} \rho \Delta - \frac{a c^{2}_{s}}{1 + w} D^{2} \Delta \nonumber \\
& + &
\frac{\kappa^{2}_{_{5}} \psi}{1 + w}  c^{2}_{s} \Delta - a^{2} \kappa^{2}_{_{5}} D^{2} \psi
\end{eqnarray}

Obviously, the equations are too complicated and it is almost impossible to have a possible solution
from these complicated equations {\em a priori}. However, the equations turn out to be tractable 
if we incorporate certain simplifications following physical arguments. 
We have seen in the case of Newtonian analysis that for a  
radiative bulk the Weyl fluid, in general, evolves as
\begin{equation}
\rho^* = C_0 a^{-(4 - \alpha)}
\end{equation}
with the parameter $\alpha$ in the range $1 < \alpha <4$.
We now consider $\psi$ to be a function of time only. This essentially means that we
are considering only the time-evolution of the bulk-brane scenario, which is relevant
for perturbation analysis. We further assume that
the energy exchange between the two fluids is in equilibrium. 
  This basically describes the late time behaviour.
Once again this is consistent so far as density perturbations are concerned.
Consequently, the Weyl fluid now behaves as
\be
\rho^{*} \propto {a^{-3/2}}
\ee
with the parameter $\alpha = \frac{5}{2}$.
This readily suggests that the Weyl fluid redshifts more slowly than ordinary matter
and hence, can dominate over matter at late times, reflecting one of the fundamental 
properties of dark matter.
This also provides a more stringent bound for the value of $\alpha$ from theoretical ground alone.
Later, we shall confront this value with observations.

With the above results, 
the evolution equation for the Weyl fluid at late times is radically
simplified by using $\Delta^{(b)} << \Delta^{*}$, which can be recast
in convenient form as 
\begin{equation}
 \ddot\Delta^{*} + \frac{A}{t} \dot\Delta^{*} - \left( \frac{{B}}{t} +\frac{{C}}{t^2} \right) \Delta^{*} = 0
\end{equation}
where the constants $A, B, C$ are readily determinable.

The above equation for $\Delta^{*}$ turns out to be somewhat tractable. 
One of its solutions is given by \cite{suden}
\begin{equation} 
 \Delta^{*} \sim t^{\frac{1}{2}-\frac{A}{2}} {\rm Bessel}I \left[ \sqrt{1 - 2A + A^2 + 4C}, 2 \sqrt{B} \sqrt{t} \right] 
 \end{equation}

We know that the Bessel function is a growing function.
Therefore, the evolution equation for the Weyl fluid, indeed, shows a growing mode solution,
which is required to explain the growth of perturbations at late times.

Thus, even the relativistic perturbations show that so far as theoretical results are
concerned, the Weyl fluid can explain structure formation.


\section{Confrontation with observations}

An important issue is to confront this theoretical model
with observations.
Recently there has been some study in this direction \cite{obs}, although an extensive study 
still remains as an open issue. In this section, we shall very briefly mention
what has been done in this direction so far. 

In terms of dimensionless parameters
\be
 \Omega _{\Lambda }=\frac{\Lambda }{3H_{0}^{2}},
\Omega _{\rho } =\frac{\kappa ^{2}\rho_{0}}{3H_{0}^{2}},
\Omega _{*}=\frac{2C_{0}}{a_{0}^{4-\alpha}H_{0}^{2}},
\Omega _{\lambda} =\frac{\kappa ^{2}\rho _{0}^{2}}{6\lambda H_{0}^{2}}
\ee
with the total density satisfying the critical value
\be
\Omega _{\rm tot} =\sum_{i} \Omega_i =1
\ee
the Friedmann equation (including the contribution from brane tension)
can be re-written as
\be
\frac{H^{2}}{H_{0}^{2}}=\Omega _{\Lambda }+\Omega _{\rho}\frac{{a_{0}^{3}}}{{a^{3}}}
+\Omega_{*}\frac{{a_{0}^{4-\alpha }}}{{a^{4-\alpha }}}
+\Omega _{\lambda}\frac{{a_{0}^{6}}}{{a^{6}}}
\ee
The luminosity distance for FRW branes is now given by
\begin{center}
 $d_{L}\left( z\right) =\frac{\left( 1+z\right) a_{0}}{H_{0}} \int_{a}^{a_{0}}\frac{ada}{\left[ \Omega _{\Lambda }a^{6}
+\Omega _{\rho }a_{0}^{3}a^{3}+\Omega_{*}a_{0}^{4-\alpha }a^{\alpha +2}+\Omega _{\lambda }a_{0}^{6}\right] ^{1/2}}$
\end{center}
For $\Omega_\lambda \rightarrow 0$ relevant for late-time cosmology, this integral
can be evaluated as
\be
d_{L}^{\Lambda\lambda *}=d_{L}^{\Lambda {\rm CDM}}+\Omega _{*}I_{*}
\ee
where $I_{*}$ is a function having elliptic integrals of 1st and 2nd kind,
which have exact analytical expressions.

Comparing this with standard $\Lambda$CDM scenario, one finds that a certain amount of
Weyl fluid with $2 \leq \alpha \leq 3$ is in nice agreement with SNe data \cite{obs}.

It is important to mention here that in the relativistic analysis we have found that
 $\alpha = \frac{5}{2}$, which falls in this region, and hence, the braneworld model
of perturbations is so far an observationally accepted model. 
However, as already mentioned,
an extensive study in this direction remains as a very crucial open issue,
which, we hope, will be addressed in details in near future.


\section{Summary and outlook}

We have discussed certain features of cosmology in a generalised RS II braneworld scenario,
where the bulk is either a Schwarzschild-anti de Sitter or a  radiative Vaidya-anti de Sitter black hole. We have shown that the theory leads to a modified version of the brane Friedmann equations.  Specifically, the local corrections to the Friedmann equations are manifest via a quadratic contribution from the brane perfect fluid whereas the nonlocal  corrections supply a Weyl fluid which arises as an effect of the bulk-brane geometry. 
We have investigated for the role of each of the terms for early times as well as for late time cosmologies.
Further, we have shown that the Weyl fluid plays a crucial role in late time cosmology,
for the most general bulk metric. We have demonstrated that its presence radically changes the perturbation equations, and have discussed some of the implications of fluctuations involving it. The fluctuations are found to  grow at late times and thus, may take care of the  large amount of inhomogeneities observed in the local universe. We also  mentioned some observable sides 
of this model.

Some  future directions related to  branewolrd cosmology involve
both theoretical and observational aspects. In the theoretical side, 
a thorough study of different parameters related to cosmological perturbation
need to be performed. In the observational side,
confronting this model with observations in more details will test the theory in a 
more conclusive way. 
Last, but not the least, the expansion history of the universe from these
modified equations can be progressed further, which may lead to
more interesting results and may even be investigated to find out any possible link with observations.


\end{document}